\documentclass[showpacs,preprintnumbers,amsmath,amssymb,12pt]{article}
\usepackage{graphicx,amsfonts,amssymb,amsthm,amsmath,graphics,epsfig,pstricks,fancyhdr,fancybox}
\usepackage{bm}

\newcommand{\refeq}[1]{~(\ref{#1})}
\newcommand{\myref}[1]{~\ref{#1}}
\newcommand{\mycite}[1]{~\cite{#1}}

\newcommand{\omp}{\omega_\perp}

\begin{document}
\thispagestyle{empty}

\title{\Huge \textbf{The falling pencil:\\
a \emph{Divertimento} in four movements}}

\author{
Nicola \textsc{Cufaro Petroni}\footnote{cufaro@ba.infn.it}  \\
Dipartimento di \textsl{Matematica} and \textsl{TIRES}, University
of Bari (\textit{Ret})\\
 \vspace{7pt}
and \textsl{INFN} Sezione di Bari; via E. Orabona 4, 70125 Bari,
Italy}

\date{}

\maketitle

\vspace{-1cm}

\begin{abstract}
\noindent The dynamics of a simple pencil with a tip laid on a rough
table and set free to fall under the action of gravity is
scrutinized as a pedagogic case study. The full inquiry is
anticipated by a review of three other simplified movements
foreshadowing its main features. A few exact and general results
about the sliding angles and the critical static coefficient of
friction are established
\end{abstract}


\section{Prelude}
\label{Intro}

When he was a beginner in his physics studies the author of these
lines was not very adroit in solving exercises.  That
notwithstanding he managed to pass his exams and he subsequently
acquired the usual skills -- and even some zest -- in designing and
answering problems: this was of course also a result of his first
acquaintance with the teaching. In those years he posed to himself
some seemingly simple questions that he could not immediately answer
and that he did not happen to find discussed on his handbooks; but
then he dropped them and went along his way without caring too much,
even if every now and again they popped up in his head. He remembers
in particular asking himself what exactly happens to a simple pencil
with a tip laid on a table and set free to fall under the action of
gravity: would the tip on the table stay put at its initial
position, or will it begin to slide, and when? And what is its
subsequent movement? The author didn't spend in fact too much effort
on that, and he eventually gave up, but for some unrelated reason
this query resurfaced recently in his thoughts and now -- being
today retired -- he decided to devote some time in finding an
elementary, but satisfactory answer: a pursuit prompted by sheer
curiosity and to him comparable to a \emph{Divertimento} that
hopefully could also be of some interest for students and scholars

In order to tackle this case study in a pedagogic style the
discussion has been articulated in four sections corresponding to
different possible \emph{movements} of growing difficulty: in the
first one (Section\myref{rotating}) the pencil tip is hinged in a
point and the system is free to rotate without friction around it
sweeping an arbitrary fall angle $0\le\theta\le2\pi$ (in this
section there is no table to speak about). This simplified setting
will lend the possibility of studying the hinge reaction forces
without making any reference to the friction. This smoothness
requirement is carried on also in the two subsequent sections where
the second and third movement are investigated: in the
Section\myref{rail} the pen tip is restrained to slide along a
horizontal frictionless rail (here again $\theta$ is allowed to go
from $0$ to $2\pi$) so that a first idea of what happens in this
limiting case is acquired. Then in the Section\myref{step} the
horizontal table appears (so that now $0\le\theta\le\,^\pi/_2$): it
is still frictionless, but featuring a step that forbids an early
sliding of the pencil on one side. This third movement allows to
recognize that beyond an angle
$\theta_r=\arccos\,^2/_3\simeq0.268\,\pi$ the pencil tip begins to
slide on the step-free side. In the Section\myref{rough} we finally
turn our attention to the fourth movement of the free pencil on a
rough table where $\mu_s$ and $\mu_\kappa$ respectively are the
static and kinetic coefficients of friction. In this case it is
found that there is a precise critical value
$\overline{\mu}_s=2^{-\,^{13}/_2}\cdot3\cdot5^{\,^3/_2}\simeq0.371$
of the static coefficient beyond which no early sliding is allowed
(much as if the step of the third movement was in place). A further
fallout of this finding is that there are exact angles
\begin{equation*}
    \overline{\theta}=\arccos\frac{9}{11}\simeq0.195\,\pi
    \qquad\qquad\overline{\overline{\theta}}=\arccos\frac{48\sqrt{14}-35}{231}\simeq0.285\,\pi
\end{equation*}
such that an early sliding (for $\mu_s\le\overline{\mu}_s$) can
happen only at $\theta^*\le\overline{\theta}$, while a later sliding
on the opposite side (for $\mu_s\ge\overline{\mu}_s$) only starts at
$\theta^{**}\ge\overline{\overline{\theta}}$. It is worthwhile to
remark that the values of
$\theta_r,\overline{\mu}_s,\overline{\theta}$ and
$\overline{\overline{\theta}}$ are universal for every idealized bar
used as a pencil and for every kind of rough table used to perform
the experiment. The values either of $\theta^*$ or of $\theta^{**}$
on the other hand apparently depend on $\mu_s$. The trajectories of
the center of mass of the pencil for the third and fourth movement
are also investigated, those for the first and second movement being
utterly trivial. A few final remarks are ultimately added in the
last Section\myref{concl}

\section{First movement: The hinge}\label{rotating}

\begin{figure}
\begin{center}
\includegraphics[width=8cm]{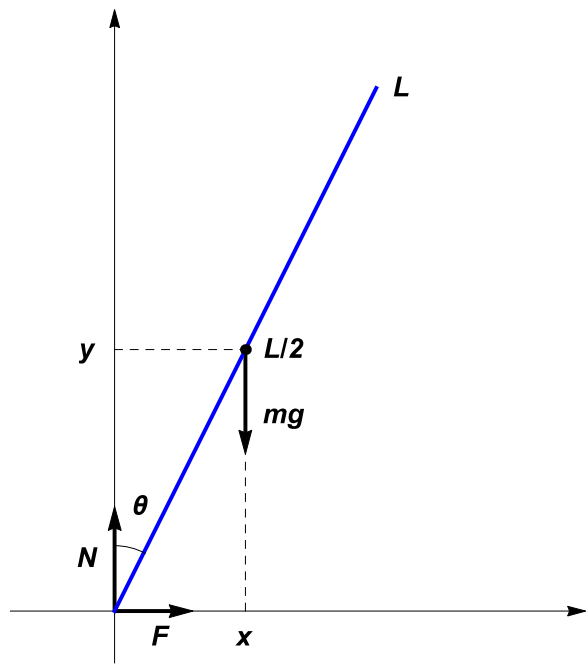}
\caption{The falling pencil.}\label{pencil}
\end{center}
\end{figure}

Consider a homogeneous, rigid rod (the \emph{pencil}) of mass $m$
and length $L$ with one of its extremities in contact with a
horizontal surface ((the \emph{table}) and suppose that $\mu_s$ and
$\mu_\kappa$ respectively are the static and kinetic friction
coefficients (see Figure\myref{pencil}). Let $\theta$ be the angle
between the pencil and the vertical to the surface, and $x,y$ the
coordinates of the middle point (the center of mass, \emph{CM}) in a
plan containing the pencil and the vertical so that (when the pencil
tip stays still in the axes origin)
\begin{equation}\label{coord}
    x=\frac{L}{2}\,\sin\theta\qquad\quad
    y=\frac{L}{2}\,\cos\theta\qquad\qquad0\le\theta\le\,^\pi/_2
\end{equation}
We will denote in the following as $ N$ and $ F$ respectively the
vertical and horizontal components of the ground reaction force:
apparently $F$ is non-zero only if a friction is there. The aim of
the present paper is a discussion of the dynamics of the falling
pencil, and in particular of its behavior when it also possibly
slips on the surface before touching the ground

We will suppose for simplicity at first that the pencil is not
allowed to move along the surface: for instance we can imagine it
hinged at the axes origin and free to rotate without friction around
it. We will also admit that it can go full circle -- as if the table
were not there -- so that now $0\le\theta\le2\pi$. This would enable
us to study the reaction forces $ N$ and $ F$ in detail in an
initially simplified setting that will be useful in the subsequent
discussion. We have indeed in this case just a \emph{physical
pendulum} (an extended rigid body) performing swings of arbitrary
amplitude. The topic is very well known and has been widely studied,
for instance as \emph{inverted pendulum} w.r.t.\ the stabilization
of its equilibrium (see for instance\mycite{frank},\mycite{liber}
and\mycite{www}): we will however skip these topics altogether by
confining ourselves just to a simplified discussion of the circular
pendulum.

The Newton equations of motion, with a fixed point in the origin,
can be simply written in this case as
\begin{equation}\label{newt}
    m\ddot{x}=F\qquad\quad m\ddot{y}=N-mg\quad\qquad
    I_0\ddot{\theta}=mgx=mg\frac{L}{2}\sin\theta
\end{equation}
where $I_0=\,^{mL^2}/_3$ is the moment of inertia of the pencil
w.r.t.\ its fixed end. Neglecting for the time being the first two
equations, we focus our attention on the third that can be written
as
\begin{equation}\label{thetaeq}
    \ddot{\theta}=\frac{\omp^2}{2}\,\sin\theta\qquad\quad\omp=\sqrt{\frac{3g}{L}}
\end{equation}
There is not an explicit elementary solution of this non linear
equation, but that notwithstanding we can study it in some detail.
It is easy to see indeed that
\begin{equation*}
    \frac{d}{dt}\big(\dot{\theta}^2\big)=2\dot{\theta}\ddot{\theta}=\omp^2\,\dot{\theta}\sin\theta=-\omp^2\frac{d}{dt}\left(\cos\theta\right)
\end{equation*}
and therefore
\begin{equation}\label{intc}
    \dot{\theta}^2=-\omp^2\,\cos\theta+c
\end{equation}
where $c$ is an arbitrary integration constant depending on the
initial conditions. Let us make at first (a bit naively) what seems
to be the simplest choice, namely
\begin{equation}\label{initial0}
    \theta(0)=0\qquad\qquad\dot{\theta}(0)=0
\end{equation}
In this case apparently we have $c=\omp^2$ and hence
\begin{equation*}
    \dot{\theta}^2=\omp^2(1-\cos\theta)\ge0\qquad\quad0\le\theta\le2\pi
\end{equation*}
or in another form
\begin{equation*}
    \omega(\theta)=\frac{d\theta}{dt}=\omp\sqrt{1-\cos\theta}
\end{equation*}
This non-linear, first order equation -- which also shows that
$\omp$ is the angular velocity at $\theta=\,^\pi/_2$ -- can be
easily solved by separating the variables, namely
\begin{equation*}
    \int_0^\theta\frac{d\phi}{\sqrt{1-\cos\phi}}=\omp\int_0^tds=\omp t
\end{equation*}
but it can be seen that the left hand integral diverges because the
integrand function has a non integrable singularity in the origin:
\begin{equation*}
    \frac{1}{\sqrt{1-\cos\phi}}=O\big(\phi^{-1}\big)\qquad\quad\phi\to0
\end{equation*}
We have indeed from L'H\^opital rule that
\begin{figure}
\begin{center}
\includegraphics[width=11cm]{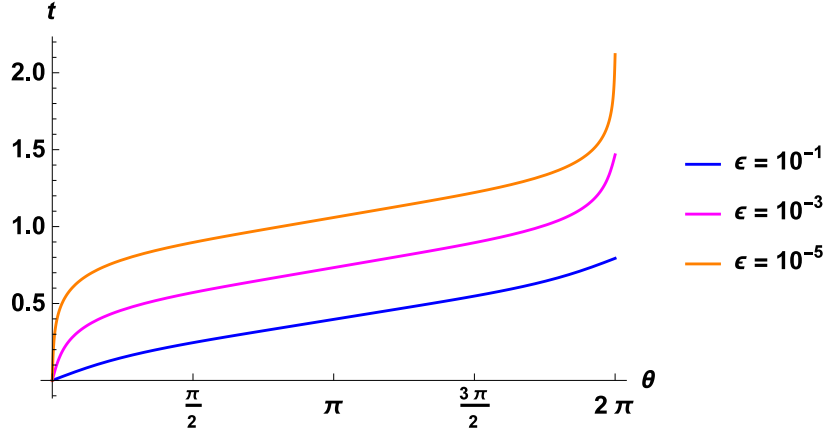}
\caption{The time $t$ (in seconds) needed to reach an angle $\theta$
according to\refeq{implicit} for three different values of
$\epsilon$ and $\omp=10\;\text{{sec}}^{-1}$ (corresponding to an $L$
of roughly $30$ cm). The pencil is allowed to go full circle from
$0$ to $2\pi$.}\label{ttheta}
\end{center}
\end{figure}
\begin{equation*}
   \lim_{\phi\to0}\frac{\phi}{\sqrt{1-\cos\phi}}=2\lim_{\phi\to0}\frac{\sqrt{1-\cos\phi}}{\sin\phi}
   =2\lim_{\phi\to0}\sqrt{\frac{1-\cos\phi}{1-\cos^2\phi}}=2\lim_{\phi\to0}\frac{1}{\sqrt{1+\cos\phi}}=\sqrt{2}
\end{equation*}
As a matter of fact this behavior is rather understandable and can
be traced back to our awkward choice of the initial conditions: when
indeed we assume\refeq{initial0} we are putting the system in its
position of unstable equilibrium, and therefore the pencil would
ideally stand up forever so that the time needed to reach a position
$\theta\neq0$ would diverge. We need therefore to take a slightly
different (and more realistic) initial condition, for instance with
a gentle push onward
\begin{equation}\label{initial}
    \theta(0)=0\qquad\qquad\dot{\theta}(0)=\omega_0>0
\end{equation}
where $\omega_0$ can be chosen small and even infinitesimal to
approach the ideal (but singular) condition\refeq{initial0}. With
this new assumption the integration constant in\refeq{intc} becomes
$c=\omega_o^2+\omp^2$ and the equation takes the form
\begin{equation}\label{angvel}
    \dot{\theta}^2=\omega_0^2+\omp^2(1-\cos\theta)
\end{equation}
to wit
\begin{equation*}
    \omega(\theta)=\frac{d\theta}{dt}=\omp\sqrt{2\epsilon+1-\cos\theta}\qquad
    \qquad2\epsilon=\frac{\omega_0^2}{\omp^2}
\end{equation*}
This equation can be solved again by separating the variables
\begin{equation}\label{theta}
    \int_0^\theta\frac{d\phi}{\sqrt{1+2\epsilon-\cos\phi}}=\omp\int_0^tds=\omp t
\end{equation}
but now (see\mycite{grad} 2.571.5) the left hand side integral
converges for $\epsilon>0$ and we have
\begin{equation}\label{implicit}
    \sqrt{\frac{2}{1+\epsilon}}F\left(\arcsin\sqrt{(1+\epsilon)\frac{1-\cos\theta}{1+2\epsilon-\cos\theta}},\,\sqrt{\frac{1}{1+\epsilon}}\right)=\omp t
\end{equation}
where (see\mycite{grad} 8.111.2)
\begin{equation*}
    F(\varphi,b)=\int_0^\varphi\frac{d\alpha}{\sqrt{1-b^2\sin^2\alpha}}\qquad\quad
    b^2<1
\end{equation*}
is the elliptic integral of the first kind. As a matter of fact the
equation\refeq{implicit} gives the function $\theta(t)$ in an
implicit form that is not easy to invert, and this form moreover is
not much more manageable than the original integral
formulation\refeq{theta} because the function $F(\varphi,b)$ is
nothing else than a name for another integral. Since however these
integrals are nowadays numerically performed by the usual
mathematical software, the results\refeq{theta} and\refeq{implicit}
can easily be used to plot the function $t(\theta)$, \emph{time
needed to reach an angle $\theta$}, as in the Figure\myref{ttheta}
where, with an exchange of the coordinate axes, we would also get a
graphical representation of $\theta(t)$

\begin{figure}
\begin{center}
\includegraphics[width=11cm]{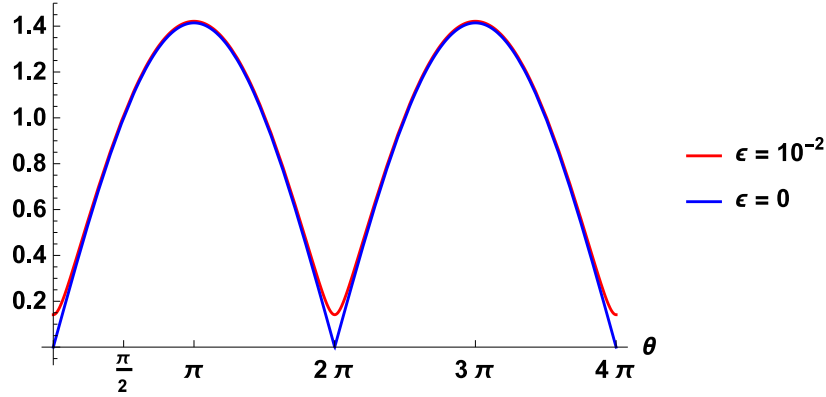}
\caption{The dimensionless angular velocity $^\omega/_{\omp}$
of\refeq{omega} as a function of $\theta$ for two different values
of $\epsilon$.}\label{om}
\end{center}
\end{figure}

We can next take advantage of the first two equations\refeq{newt} to
find the reaction forces $N$ and $F$: since from\refeq{coord} it is
\begin{equation}\label{accel}
  \ddot{x}=\frac{L}{2}\left(\ddot{\theta}\cos\theta-\dot{\theta}^2\sin\theta\right)
  \qquad\quad
  \ddot{y}=-\frac{L}{2}\left(\ddot{\theta}\sin\theta+\dot{\theta}^2\cos\theta\right)
\end{equation}
from the first two equations in\refeq{newt} we have
\begin{equation*}
  F=\frac{mL}{2}\left(\ddot{\theta}\cos\theta-\dot{\theta}^2\sin\theta\right)
  \qquad\quad
  N=mg-\frac{mL}{2}\left(\ddot{\theta}\sin\theta+\dot{\theta}^2\cos\theta\right)
\end{equation*}
and, since we know that, with the initial conditions\refeq{initial},
the equations\refeq{thetaeq} and\refeq{angvel} hold, after a little
algebra we find how the reaction forces vary as functions of
$\theta$
\begin{figure}
\begin{center}
\includegraphics[width=11cm]{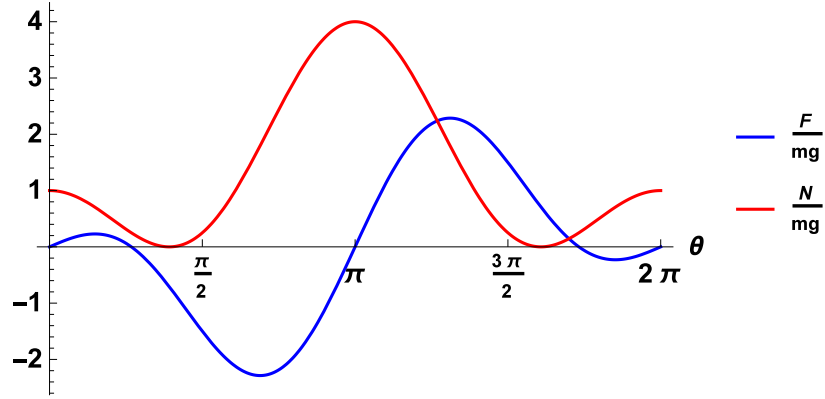}
\caption{The dimensionless reaction forces $\,^F/_{mg}$ and
$\,^N/_{mg}$ of\refeq{F0} and\refeq{N0} (for $\epsilon=0$) as
functions of the position $\theta$.}\label{reactions}
\end{center}
\end{figure}
\begin{eqnarray}
    F(\theta)&=&\frac{3\,mg}{2}\left(\frac{3}{2}\cos\theta-1-2\epsilon\right)\sin\theta\label{F}\\
    N(\theta)&=&\frac{mg}{4}+\frac{3\,mg}{2}\left(\frac{3}{2}\cos\theta-1-2\epsilon\right)\cos\theta\label{N}
\end{eqnarray}
From\refeq{angvel} moreover it is also possible to show that the
angular velocity $\omega=\dot{\theta}$ varies with the position
$\theta$ according to the formula
\begin{equation}\label{omega}
    \omega(\theta)=\omp\sqrt{2\epsilon+1-\cos\theta}
\end{equation}
It is interesting to remark at this point that, while the time
formula\refeq{theta} is singular for $\epsilon\to0^+$, the
equations\refeq{F},\refeq{N} and\refeq{omega} continuously go into
their $\epsilon=0$ forms
\begin{eqnarray}
    F(\theta)&=&\frac{3\,mg}{2}\left(\frac{3}{2}\cos\theta-1\right)\sin\theta\label{F0}\\
    N(\theta)&=&\frac{mg}{4}+\frac{3\,mg}{2}\left(\frac{3}{2}\cos\theta-1\right)\cos\theta\label{N0}\\
    \omega(\theta)&=&\omp\sqrt{1-\cos\theta}\label{omega0}
\end{eqnarray}
corresponding to the null initial conditions\refeq{initial0}: these
limiting formulas can now be properly used to represent the simplest
behavior of the reaction forces and of the angular velocity at every
possible position $\theta$. In the Figure\myref{reactions} we have
plotted the dimensionless functions $\,^F/_{mg}$ and $\,^N/_{mg}$
of\refeq{F0} and\refeq{N0} (with $\epsilon=0$), while the
velocity\refeq{omega} of\refeq{omega0} in its dimensionless form
$\,^\omega/_{\omp}$ is plotted in the Figure\myref{om} in the
interval $[0,4\pi]$: for $\epsilon>0$, $\omega(\theta)$ turns out to
be a smooth function even at the angles $\theta=0,2\pi,4\pi\ldots$
Apparently when we also plug into these formulas the function
$\theta(t)$ implicitly defined in\refeq{implicit} we also get the
time dependence of $F,N$ and $\omega$, but, needless to say, this
would be a cumbersome task that we will neglect here

It is worthwhile to remark finally that the two reaction components
$N,$ and $F$ also take negative values: for $F$ this is apparent
from the Figure\myref{reactions} and we see from\refeq{F0} that --
even with $\epsilon=0$ and remaining just in the interval
$[0,\,^\pi/_2]$ -- we have $F<0$ provided that
\begin{equation*}
    \frac{2}{3}>\cos\theta\ge0\qquad
    \arccos\left(\,^2/_3\right)<\theta\le\frac{\pi}{2}\qquad
    \arccos\left(\,^2/_3\right)\simeq0.268\,\pi
\end{equation*}
As for the normal component we find instead from\refeq{N} that $N$
can be negative only if $\epsilon>0$: more precisely we have $N\le0$
when
\begin{equation*}
    \frac{1}{3}+\frac{2}{3}\left(\epsilon+\sqrt{\epsilon(1+\epsilon)}\right)\ge\cos\theta\ge\frac{1}{3}+\frac{2}{3}\left(\epsilon-\sqrt{\epsilon(1+\epsilon)}\right)
\end{equation*}
namely for $\theta$ falling in an interval that shrinks to the
single point $\arccos(\,^1/_3)\simeq0.392\,\pi$ for $\epsilon=0^+$.
These negative values will be of some consequence in the sequel
because they will suggest where an \emph{un}-hinged pencil will
begin a sliding movement when the available constraints will be
unable to provide a \emph{negative} reaction
\begin{figure}
\begin{center}
\includegraphics[width=8cm]{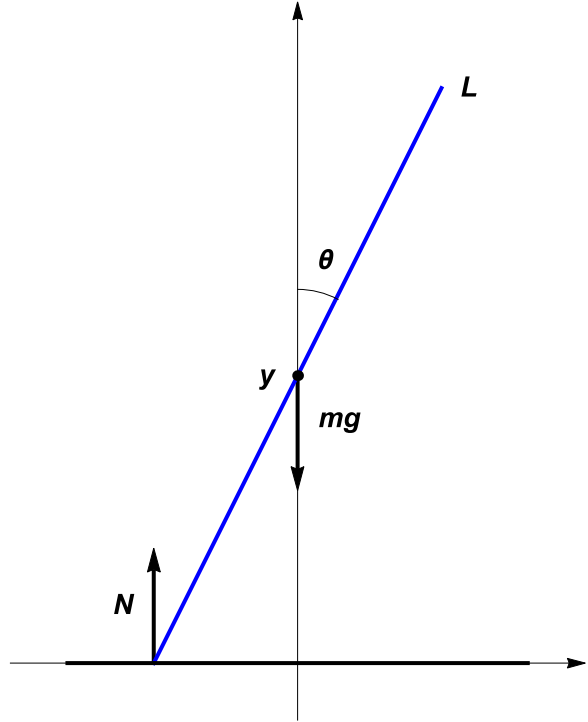}
\caption{The pencil tip sliding along a frictionless
rail.}\label{pencil2}
\end{center}
\end{figure}

\section{Second movement: The rail}\label{rail}

Before going ahead to our pencil with one end laid on a horizontal
rough table and free to move along it, we will stop for a while to
consider two more frictionless cases. In order to allow again for a
full swing of the system from $0$ to $2\pi$, moreover, in the first
of these examples we will suppose that the pencil tip on the $x$
axis in the Figure\myref{pencil2} is in fact constrained to slide
along a rail without leaving it while the center of mass goes from
$\,^L/_2$ to $\,^{-L}/_2$ and back again. At variance with the case
of the previous section, however, now there is no horizontal force
$F$ because neither friction nor hinges in the axes origin are
present. As a consequence the pencil \emph{CM} will simply move
along the $y$ axis with $x=0$, if its movement starts with this
initial condition. On the other hand there is no longer a fixed
point of the system so that now its rotational dynamics is better
accounted for by looking at its motion around the \emph{CM}.
Therefore the Newton equations now are
\begin{figure}
\begin{center}
\includegraphics[width=11cm]{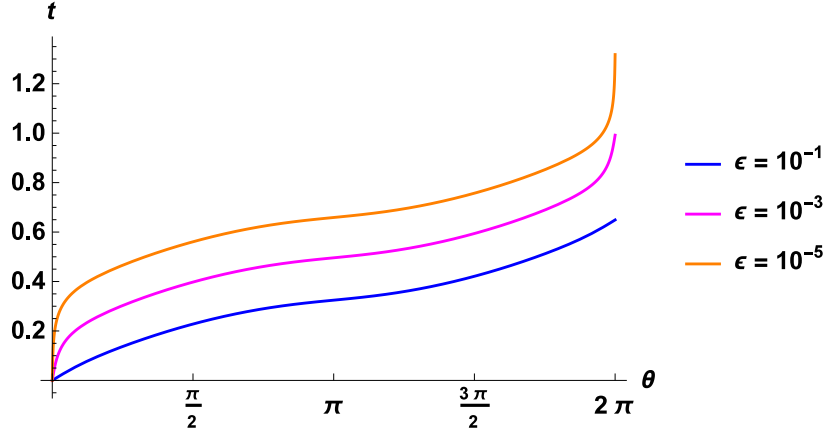}
\caption{The time $t$ (in seconds) needed to reach an angle $\theta$
according to\refeq{theta2} for three different values of $\epsilon$
and $\omp=10\;\text{{sec}}^{-1}$. The pencil is allowed to go full
circle from $0$ to $2\pi$.}\label{ttheta2}
\end{center}
\end{figure}
\begin{equation}\label{newt2}
    m\ddot{x}=0\qquad\quad m\ddot{y}=N-mg\quad\qquad
    I_{CM}\ddot{\theta}=N\frac{L}{2}\sin\theta
\end{equation}
where $I_{CM}=\,^{mL^2}/_{12}$ is the moment of inertia of the
pencil w.r.t.\ its \emph{CM}, while the geometrical relations among
the coordinates become
\begin{equation}\label{coord2}
    x=0\qquad\qquad
    y=\frac{L}{2}\,\cos\theta
\end{equation}
To tackle our problem we can now retrace a path similar to that
followed in the Section\myref{rotating}: from\refeq{newt2} the
rotational acceleration around the \emph{CM} is
\begin{equation}\label{accel2}
    \ddot{\theta}=\frac{6N}{mL}\sin\theta
\end{equation}
while on the other hand again from\refeq{newt2} and
from\refeq{accel} we find
\begin{equation}\label{reactN}
    N=m\left(\ddot{y}+g\right)=mg-\frac{mL}{2}\left(\ddot{\theta}\sin\theta+\dot{\theta}^2\cos\theta\right)
\end{equation}
so that altogether it is
\begin{equation*}
    \ddot{\theta}=\sin\theta\left[2\omp^2-3\left(\ddot{\theta}\sin\theta+\dot{\theta}^2\cos\theta\right)\right]
\end{equation*}
It is easy to see now that
\begin{eqnarray*}
  \frac{d}{dt}\big(\dot{\theta}^2\big)
  &=&2\dot{\theta}\ddot{\theta}\;=\;2\dot{\theta}\sin\theta\left[2\omp^2-3\left(\ddot{\theta}\sin\theta+\dot{\theta}^2\cos\theta\right)\right] \\
   &=&-\frac{d}{dt}\left[4\omp^2\cos\theta+3\left(\dot{\theta}\sin\theta\right)^2\right]
\end{eqnarray*}
to wit
\begin{equation}\label{firstint}
    \dot{\theta}^2\left(1+3\sin^2\theta\right)+4\omp^2\cos\theta=c
\end{equation}
and since with the slightly off-equilibrium initial
conditions\refeq{initial}, and keeping the same notations, it is
\begin{equation*}
    c=4\omp^2+\omega_o^2=4\omp^2\left(1+\frac{\epsilon}{2}\right)\qquad\qquad\epsilon=\frac{\omega_0^2}{2\omp^2}
\end{equation*}
we finally have
\begin{figure}
\begin{center}
\includegraphics[width=11cm]{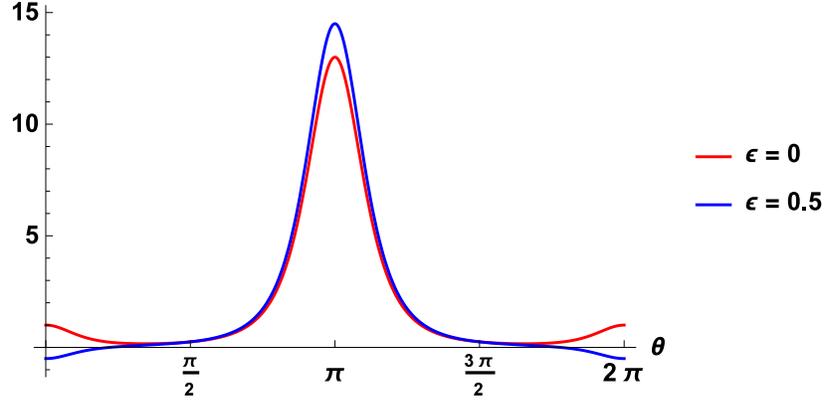}
\caption{The dimensionless reaction force $\,^N/_{mg}$ of\refeq{N2}
and\refeq{N20} as a function of the position $\theta$ for two values
of $\epsilon$.}\label{reactions2}
\end{center}
\end{figure}
\begin{equation}\label{angvel2}
    \dot{\theta}^2=\omp^2\,\frac{2\epsilon+4(1-\cos\theta)}{1+3\sin^2\theta}=\omp^2\,\frac{2\epsilon+4(1-\cos\theta)}{4-3\cos^2\theta}
\end{equation}
This equation can be integrated again by separating the variables
giving
\begin{equation}\label{theta2}
    \int_0^\theta\sqrt{\frac{4-3\cos^2\phi}{2\epsilon+4(1-\cos\phi)}}\,d\phi=\omp t=t\sqrt{\frac{3g}{L}}
\end{equation}
and while this implicit solution has no elementary inverse function
it is possible to numerically evaluate the integral to calculate the
time $t$ needed to reach an angle $\theta$: the results plotted in
the Figure\myref{ttheta2} show a qualitative behavior similar to
that of the Figure\myref{ttheta}. Here too, however, the time $t$
diverges when $\epsilon\to0^+$

To study next the reaction force $N$ we plug\refeq{accel2}
and\refeq{angvel2} into\refeq{reactN} obtaining the equation
\begin{equation*}
    N=mg-\frac{mL}{2}\left(\frac{6N}{mL}\sin^2\theta+\frac{3g}{L}\,\frac{2\epsilon+4(1-\cos\theta)}{1+3\sin^2\theta}\cos\theta\right)
\end{equation*}
that is easily solved providing
\begin{equation}\label{N2}
    N=mg\,\frac{4+3\cos^2\theta-3(\epsilon+2)\cos\theta}{(4-3\cos^2\theta)^2}
\end{equation}
For $\epsilon=0$ this simply becomes
\begin{equation}\label{N20}
    N=mg\,\frac{4+3\cos^2\theta-6\cos\theta}{(4-3\cos^2\theta)^2}
\end{equation}
The dimensionless function $\,^N/_{mg}$ is plotted in the
Figure\myref{reactions2} for two different initial conditions
$\epsilon$, and it is interesting to remark that now -- at variance
with the case discussed in the Section\myref{rotating} -- its values
are always positive, and that to have also negative values the
initial angular velocity $\omega_0$ should in fact exceed a fairly
large threshold. More precisely it would be possible to see that the
Mexican-hat shaped red curve of the Figure\myref{reactions2} bends
its tails under the $x$-axis only for $\epsilon>\,^1/_3$, namely for
$\omega_0>\sqrt{\,^{2g}/_L}\;\,\text{sec}^{-1}$: for example, for
$L=0.2\;\text{m}$, this approximately means
$\omega_0>10\;\text{sec}^{-1}$. Finally from\refeq{angvel2} we have
the angular velocity
\begin{figure}
\begin{center}
\includegraphics[width=12cm]{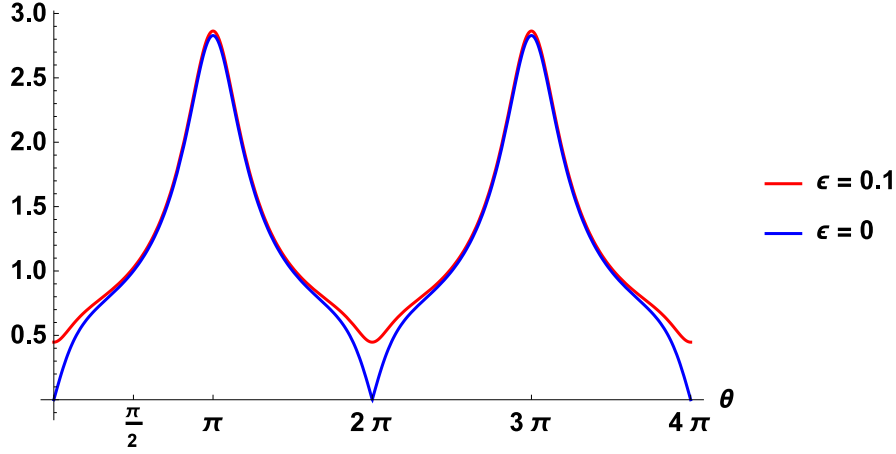}
\caption{The dimensionless angular velocities $\,^\omega/_{\omp}$
of\refeq{omega2}and\refeq{omega20} as a function of $\theta$ for two
values of $\epsilon$.}\label{om2}
\end{center}
\end{figure}
\begin{equation}\label{omega2}
    \omega(\theta)=\dot{\theta}=\omp\,\sqrt{\frac{2\epsilon+4(1-\cos\theta)}{1+3\sin^2\theta}}
\end{equation}
that is for $\epsilon=0$
\begin{equation}\label{omega20}
    \omega(\theta)=2\omp\,\sqrt{\frac{1-\cos\theta}{1+3\sin^2\theta}}
\end{equation}
In its dimensionless form $\,^\omega/_{\omp}$ this angular velocity
is reproduced in the Figure\myref{om2} for two values of $\epsilon$,
and the functions turn out to be smooth again even at
$\theta=0,2\pi,4\pi,\ldots$ for every non-zero $\epsilon>0$

\section{Third movement: The step}\label{step}

\begin{figure}
\begin{center}
\includegraphics[width=11cm]{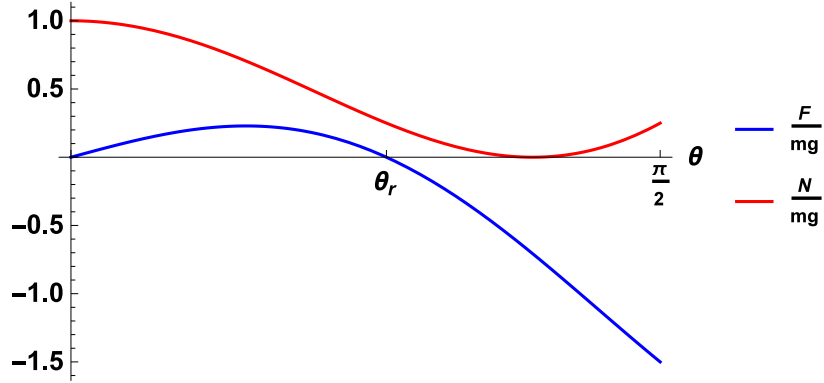}
\caption{The dimensionless reaction forces $\,^F/_{mg}$ and
$\,^N/_{mg}$\refeq{F0} and\refeq{N0} of the \emph{hinged} pencil of
the Section\myref{rotating} as functions of the position
$0\le\theta\le\,^\pi/_2$ for $\epsilon=0$. From\refeq{F0} we see
that the force $F$ reverses its sign when
$\theta\ge\theta_r=\arccos\,^2/_3$.}\label{reactions3}
\end{center}
\end{figure}

\begin{figure}
\begin{center}
\includegraphics[width=10cm]{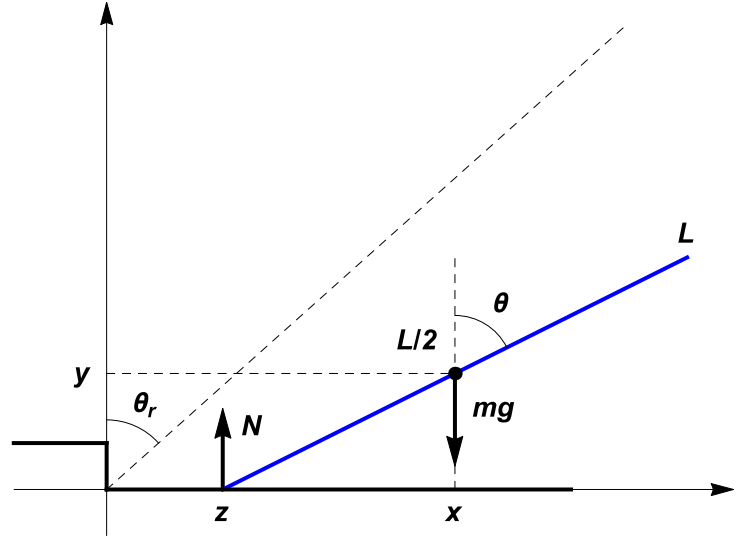}
\caption{When $\theta>\theta_r=\arccos\,^2/_3$ on a frictionless
surface with a step the pencil also starts drifting
rightwards.}\label{pencil3}
\end{center}
\end{figure}

In our third frictionless case we begin first by looking back to the
reaction forces discussed in the Section\myref{rotating}. When it
begins to fall, indeed, the hinged pencil rotates as in the
Figure\myref{pencil} with a fixed point and hence the reaction
forces vary with $\theta$ as in the Figure\myref{reactions}. To be
more precise we have reproduced in the Figure\myref{reactions3} the
forces $\,^{F(\theta)}/_{mg}$ and $\,^{N(\theta)}/_{mg}$ in the
interval  $0\le\theta\le\,^\pi/_2$ in the limiting case of
$\epsilon=0$ presented in\refeq{F0} and\refeq{N0}.
From this picture and the corresponding equations we see in
particular that, while $N$ never goes negative, the horizontal
component $F$ of the reactions in the Figure\myref{pencil} reverses
its sign beyond an angle $\theta_r=\arccos\,^2/_3\simeq0.268\,\pi$
suggesting that some force is needed to keep the pencil tip from
moving \emph{to the right} when $\theta>\theta_r$. Suppose then now
that -- without being hinged at the origin -- our pencil is just
laid on a frictionless table and allowed to fall as in the
Figure\myref{pencil}, but also that its contact tip is forbidden to
slide \emph{leftwards} (as instead it was allowed in the
Section\myref{rail}) by the presence of some obstacle, for instance
a step as in the Figure\myref{pencil3}. From the previous remarks it
follows then that when $\theta$ exceeds $\theta_r$ the pencil tip
starts sliding \emph{rightwards} because -- being now unhinged -- no
\emph{negative} horizontal reaction force can arise to prevent that.
The pencil reaches the angle $\theta_r$ at a time $t_r$ that can be
explicitly calculated from the integral\refeq{theta} for a small
initial destabilizing condition $\epsilon>0$, while at that point
its angular velocity from\refeq{omega0} is
\begin{equation*}
    \omega_r=\omega(\theta_r)=\frac{\omp}{\sqrt{3}}=\sqrt{\frac{g}{L}}
\end{equation*}
We will now investigate the movement of our system for
$\theta_r<\theta<\,^\pi/_2$ from the time $t_r$ until the instant
$T$ of the impact on the table.

If, according to the Figure\myref{pencil3}, $z$ is the position of
the contact point on the table the relationships among the variables
are now
\begin{equation}\label{coord3}
    x=z+\frac{L}{2}\,\sin\theta\qquad\quad
    y=\frac{L}{2}\,\cos\theta\qquad\qquad0\le\theta\le\,^\pi/_2
\end{equation}
while the Newton equations of motion are
\begin{equation}\label{newt3}
    m\ddot{x}=0\qquad\quad m\ddot{y}=N-mg\quad\qquad
    I_{CM}\ddot{\theta}=N\frac{L}{2}\sin\theta
\end{equation}
where again $I_{CM}=\,^{mL^2}/_{12}$. As a consequence the second
equation\refeq{accel} together with the equations\refeq{accel2}
and\refeq{reactN} still hold, and hence also\refeq{firstint} can be
deduced. Imposing then the conditions at $t=t_r$ we find that the
integration constant now is
\begin{equation*}
    c=\omega_r^2(1+3\sin^2\theta_r)+3\omp^2\cos\theta_r=\frac{32}{9}\,\omp^2
\end{equation*}
and therefore we get
\begin{equation}\label{angvel3}
    \dot{\theta}^2=\frac{4\,\omp^2}{9}\,\frac{8-9\cos\theta}{4-3\cos^2\theta}
\end{equation}
with its new corresponding time equation
\begin{equation}\label{theta3}
    \int_{\theta_r}^\theta\sqrt{\frac{4-3\cos^2\phi}{8-9\cos\phi}}\,d\phi=\frac{2\,\omp}{3}\,
    (t-t_r)
\end{equation}
that can be numerically evaluated to calculate the time $t$ needed
to reach an angle $\theta\in[\theta_r,\,^\pi/_2]$: for instance the
time $T$ when the pencil hits the floor will be
\begin{figure}
\begin{center}
\includegraphics[width=12cm]{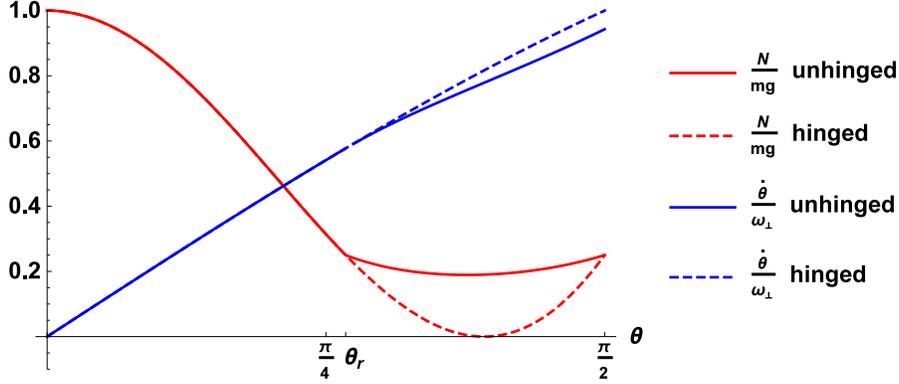}
\caption{Dimensionless reaction force $\,^N/_{mg}$ and angular
velocity $\,^{\dot{\theta}}/_{\omp}$ (continuous lines when
$\theta\ge\theta_r$) on a frictionless surface with a step, compared
with the same quantities in the case of the hinged pencil (dashed
lines) that also accounts for the movement when
$\theta<\theta_r$.}\label{Nstep}
\end{center}
\end{figure}
\begin{equation}\label{impacT}
    T=t_r+\frac{3}{2\omp}\int_{\theta_r}^{^\pi/_2}\sqrt{\frac{4-3\cos^2\phi}{8-9\cos\phi}}\,d\phi\simeq t_r+\frac{0.971}{\omp}\quad\qquad\omp=\sqrt{\frac{3g}{L}}
\end{equation}
where $t_r$ comes from\refeq{theta} choosing a small initial
condition $\epsilon>0$. As for the reaction force $N$ on the other
hand, from\refeq{accel2},\refeq{reactN} and\refeq{angvel3} we have
that
\begin{equation*}
    N=mg-\frac{mL}{2}\left(\frac{6N}{mL}\sin^2\theta+\frac{4g}{3L}\,\frac{8-9\cos\theta}{4-3\cos^2\theta}\cos\theta\right)
\end{equation*}
that eventually gives
\begin{equation}\label{N3}
    N=mg\,\frac{3\cos^2\theta-\frac{16}{3}\cos\theta+4}{(4-3\cos^2\theta)^2}
\end{equation}
The plot of $\,^N/_{mg}$ in the Figure\myref{Nstep} shows in
particular that $N$ always stays positive even in the interval
$[\theta_r,\,^\pi/_2]$ signaling that the pencil tip never leaves
the table surface. In the same Figure\myref{Nstep} also the
dimensionless angular velocity $\,^{\dot{\theta}}/_{\omp}$ is
displayed in the same interval.

To investigate next the behavior of $x,y$ and $z$ of\refeq{coord3}
we begin by remarking that the first equation in\refeq{newt3},
$m\ddot{x}=0$, clearly entails that $\dot{x}=c$ for $t\ge t_r$,
while to find the integration constant $c$ it is enough to remark
that
\begin{equation}\label{CMvx}
    \dot{x}(t)=c=\dot{x}(t_r)=\frac{L}{2}\,\omega_r\cos\theta_r=\frac{\omp
    L}{3\sqrt{3}}\qquad\quad t_r\le t\le T
\end{equation}
As a consequence we will have
\begin{equation*}
    x(t)=\frac{L}{2}\,\cos\theta_r+\frac{\omp L}{3\sqrt{3}}(t-t_r)=\frac{L}{3}+\frac{\omp L}{3\sqrt{3}}(t-t_r)\qquad\quad t_r\le t\le T
\end{equation*}
The chronological equations of $y$ and $z$, instead, can not be
deduced so simply: the second equations\refeq{newt3} for $y$, for
instance, would be nothing new w.r.t.\ the angular equation, in the
sense that if we know $\theta(t)$ we also can find $y(t)$ by taking
advantage of the second equation\refeq{coord3}. But we have seen
that the angular equation\refeq{angvel3} can not be integrated in an
elementary way, and hence even $y(t)$ has not a manageable form.
Shunning however this chronological issue, we can at least gain some
insight into the shape of the trajectory of the \emph{CM} of
coordinates $x$ and $y$.
\begin{figure}
\begin{center}
\includegraphics[width=9cm]{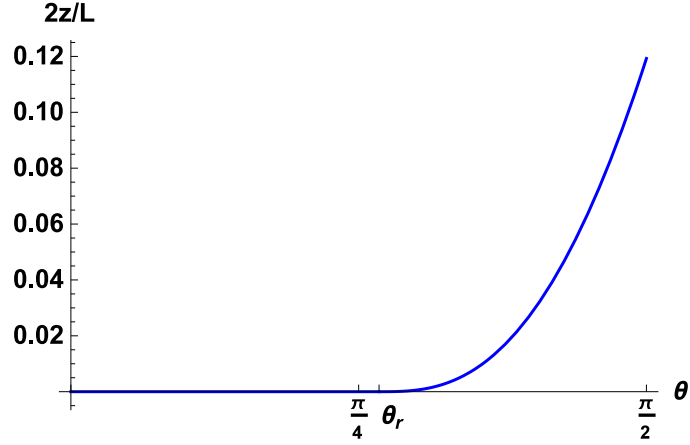}
\caption{Dimensionless position $\,^{2z(\theta)}/_L$ of the pencil
tip laid on a frictionless table with a step, as a function of
$\theta$: as long as $\theta\le\theta_r$ it is
$\,^{2z(\theta)}/_L=0$, but when $\theta\ge\theta_r$ the function
$z(\theta)=\zeta(\cos\theta)$ should be calculated
from\refeq{zeta}.}\label{z}
\end{center}
\end{figure}

It is apparent indeed that until the sliding begins (namely when
$0\le t\le t_r$ and $0\le\theta\le\theta_r$) the \emph{CM} follows a
circular path of radius $\,^L/_2$ around the origin; as soon as
$t>t_r$ and $\theta>\theta_r$, however, the \emph{CM} parts way from
the aforementioned circumference following a different flight that
can be scrutinized by looking again into the
equations\refeq{coord3}: by eliminating indeed $\cos\theta$ between
the equations
\begin{equation*}
    (x-z)^2=\frac{L^2}{4}\sin^2\theta=\frac{L^2}{4}(1-\cos^2\theta)\qquad\quad
    y^2=\frac{L^2}{4}\cos^2\theta
\end{equation*}
we have
\begin{equation*}
    y=\sqrt{\frac{L^2}{4}-(x-z)^2}
\end{equation*}
pointing to the fact that now the \emph{CM} treads along a circle,
but with a moving center in $z$. The $t$-parametric equations of
this trajectory then are
\begin{equation*}
    x(t)=z(t)+\frac{L}{2}\sqrt{1-\cos^2\theta(t)}\qquad\quad
    y(t)=\frac{L}{2}\,\cos\theta(t)
\end{equation*}
and if we define a function $\zeta(s)$ such that
\begin{equation*}
    z(t)=\zeta\big(s(t)\big)\qquad\quad s(t)=\cos\theta(t)
\end{equation*}
by adopting $s$ as a new parameter the parametric equations become
\begin{equation}\label{param}
    x(s)=\zeta(s)+\frac{L}{2}\sqrt{1-s^2}\qquad\quad
    y(s)=\frac{L}{2}\,s\qquad\quad s=\cos\theta\in[0,1]
\end{equation}
We are therefore prompted to study $\zeta(s)$:
from\refeq{coord3},\refeq{angvel3} and\refeq{CMvx} we know that
\begin{figure}
\begin{center}
\includegraphics[width=11cm]{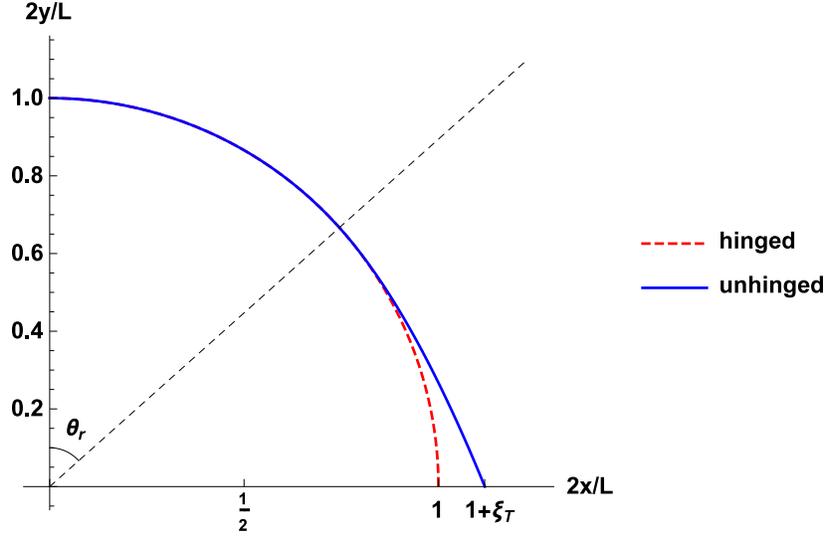}
\caption{Dimensionless \emph{CM} $xy$-trajectory: it coincides with
the circular path of the hinged pencil for $\theta\le\theta_r$, but
as soon as $\theta\ge\theta_r$ it follows a different flight with
the parametric equations\refeq{param}.}\label{landing}
\end{center}
\end{figure}
\begin{equation*}
  \dot{z} = \dot{x}-\frac{L}{2}\,\dot{\theta}\cos\theta
  =\frac{\omp L}{3\sqrt{3}}\left(1-\sqrt{3\cos^2\theta\,\frac{8-9\cos\theta}{4-3\cos^2\theta}}\right)
\end{equation*}
and since
$\dot{z}=\zeta\,'(s)\,\dot{s}=-\zeta'(s)\,\dot{\theta}\sin\theta$,
using\refeq{angvel3} again we find
\begin{equation*}
    \zeta\,'(s)=-\frac{L}{2\sqrt{3}}\left(\sqrt{\frac{4-3s^2}{(1-s^2)(8-9s)}}-\sqrt{\frac{3s^2}{1-s^2}}\right)
    \qquad\quad0\le s\le\,^2/_3
\end{equation*}
namely
\begin{equation}\label{zeta}
    \zeta(s)=\frac{L}{2\sqrt{3}}\int_s^{\,^2/_3}\left(\sqrt{\frac{4-3u^2}{(1-u^2)(8-9u)}}-\sqrt{\frac{3u^2}{1-u^2}}\right)du
    \qquad\quad0\le s\le\,^2/_3
\end{equation}
This integral, that can be performed at least numerically, lends now
the possibility of plotting both $z(\theta)=\zeta(\cos\theta)$
(Figure\myref{z}), and the trajectory parametric
equations\refeq{param} (Figure\myref{landing}) where it is
understood that $\zeta(s)=0$ when $\,^2/_3\le s\le1$. Remark that
from\refeq{zeta} we can also assess the value $x_T$ of $x$ when the
pencil finally hits the floor: if $T$, as provided by\refeq{impacT},
is the impact time, we of course have $\theta(T)=\,^\pi/_2$, namely
$s(T)=0$ and therefore the \emph{CM} $x$-coordinate when the pencil
lands on the table is
\begin{equation*}
    x_T=z_T+\frac{L}{2}=\frac{L}{2}(1+\xi_T)\qquad\quad
    z_T=z(T)=\zeta\left(s(T)\right)=\zeta(0)=\frac{L}{2}\,\xi_T
\end{equation*}
where
\begin{equation*}
    \xi_T=\frac{1}{\sqrt{3}}\int_0^{\,^2/_3}\left(\sqrt{\frac{4-3u^2}{(1-u^2)(8-9u)}}-\sqrt{\frac{3u^2}{1-u^2}}\right)du\simeq0.12
\end{equation*}
can be numerically evaluated: if for instance $L=20\;\text{cm}$,
this roughly means that $z_T\simeq1.2\;\text{cm}$
\begin{figure}
\begin{center}
\includegraphics[width=11cm]{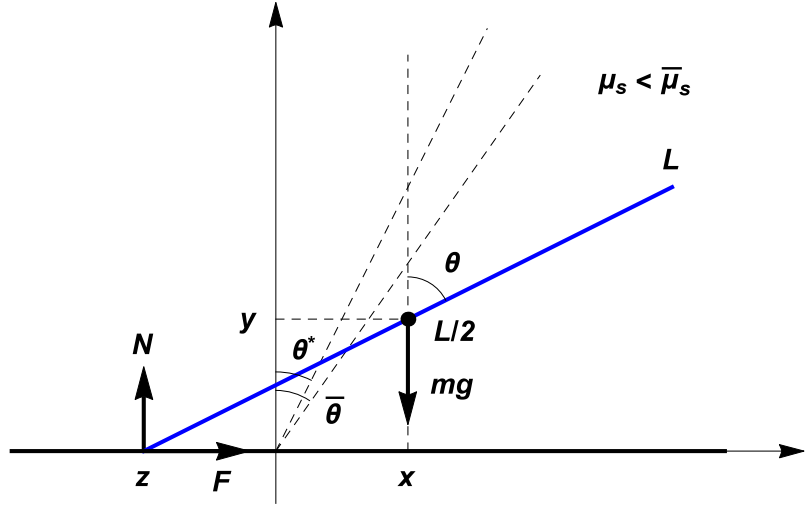}
\caption{The pencil tip on the table starts to drift leftward past
an angle
$\theta^*<\overline{\theta}=\arccos\,^9/_{11}\simeq0.195\,\pi$ if
the static friction coefficient is not large enough
($\mu_s<\overline{\mu}_s\simeq0.37$) to forbid it.}\label{pencil4}
\end{center}
\end{figure}

\section{Fourth movement: The rough surface}\label{rough}

We finally go back to our initial problem of a pencil with one end
laid on a horizontal rough table and free to slide along it while
falling, Because of the presence of friction, when the pencil starts
its movement the extremity in contact with the surface does not
move, but it can possibly slide (on both sides as we shall see) at a
later time $t^*$ when it passes beyond a position $\theta^*$: to
understand how this happens we must therefore first of all look
again at the reaction forces discussed in the
Section\myref{rotating} and\myref{step}. When it begins to fall,
indeed, the pencil rotates as in the Figure\myref{pencil} with a
fixed point and hence the reaction forces --  in the limiting case
of $\epsilon=0$ presented in\refeq{F0} and\refeq{N0} -- vary with
$\theta$ as in the Figure\myref{reactions3} with
$0\le\theta\le\,^\pi/_2$. We already remarked in the
Section\myref{step} that $F$ (the horizontal component of the
reactions in the Figure\myref{pencil}) is supposed to reverse its
sign beyond an angle $\theta_r=\arccos\,^2/_3\simeq0,268\,\pi$
(suggesting that some force is needed to keep the pencil tip from
moving \emph{to the right} when $\theta>\theta_r$), but only if it
does not start to slide \emph{to the left} at an earlier time $t^*$
at an angle $\theta^*$ as in the Figure\myref{pencil4}. This second
occurrence must indeed be taken into account because now -- in
absence of the step of the Section\myref{step} -- the static
friction force must always satisfy the condition $|F|\le\mu_sN$, and
it may happen that this requirement is not met beyond some angle
$\theta^*<\theta_r$.
\begin{figure}
\begin{center}
\includegraphics[width=12cm]{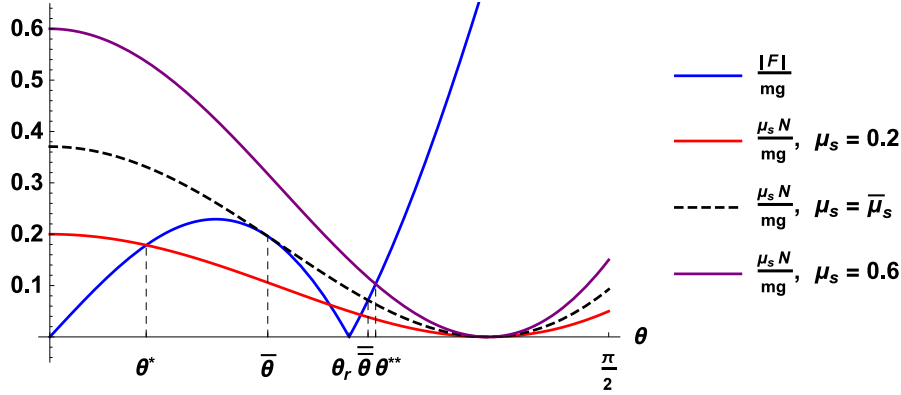}
\caption{The absolute value of the friction forces $\,^{|F|}/_{mg}$
may exceed its maximum $\,^{\mu_s N}/_{mg}$ at several possible
angles according to the value of $\mu_s$. If
$\mu_s<\overline{\mu}_s$ a slipping takes place toward the left
beyond an angle $\theta^*$; when instead $\mu_s>\overline{\mu}_s$
the pencil tip starts sliding toward the right at a later time past
the angle $\theta^{**}$.}\label{critical}
\end{center}
\end{figure}

In order to understand if and when this happens, a (dimensionless)
comparison between $|F|$ and $\mu_s N$ has been displayed in the
Figure\myref{critical} wherefrom we see that whenever $\mu_s$ is
smaller than a critical value $\overline{\mu}_s$ the pencil starts
slipping to the left at an angle $\theta^*$. From the
equation\refeq{theta} it is also possible to find the time $t^*$ of
this occurrence for every non zero initial condition $\epsilon>0$.
When instead $\mu_s>\overline{\mu}_s$, the slipping happens toward
the right at a later time $t^{**}$ when the absolute value of the
(now reversed) friction force exceeds the critical value at a larger
angle $\theta^{**}$.

To find the numerical values of these quantities we must first of
all look (with a given $\mu_s$) for the values of the angle $\theta$
such that $|F|=\mu_sN$ namely, from\refeq{F0} and\refeq{N0}, such
that
\begin{equation}\label{thetaeq1}
    \left|\frac{3}{2}\left(\frac{3}{2}\cos\theta-1\right)\sin\theta\,\right|
    =\mu_s\left(\frac{1}{4}+\frac{3}{2}\left(\frac{3}{2}\cos\theta-1\right)\cos\theta\right)
\end{equation}
when $0\le\theta\le\,^\pi/_2$. Squaring both sides and defining for
simplicity $s=\cos\theta\in[0,1]$, after a little algebra the
previous equation becomes
\begin{equation}\label{criteq}
    (1-s^2)(9s-6)^2=\mu_s^2(3s-1)^4
\end{equation}
and to search for its solutions in $[0,1]$ we recast it in the form
\begin{figure}
\begin{center}
\includegraphics[width=12cm]{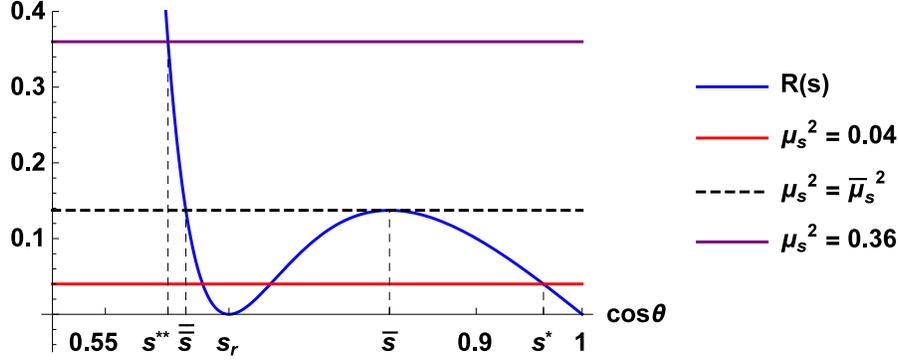}
\caption{The solutions of the equation\refeq{criteq} correspond to
the values of $s=\cos\theta$ such that $R(s)=\mu_s^2$. The critical
friction coefficient $\overline{\mu}_s$ -- beyond which no
left-slipping takes place -- coincides with the maximum of $R(s)$.
The particular values and the notation are carried over from those
of the Figure\myref{critical}.}\label{critical2}
\end{center}
\end{figure}
\begin{equation}\label{Req}
    R(s)=\frac{(1-s^2)(9s-6)^2}{(3s-1)^4}=\mu_s^2
\end{equation}
that is represented in the Figure\myref{critical2} with the same
values of $\mu_s$ adopted in the Figure\myref{critical}. It is
apparent therefrom that $s^*=\cos\theta^*$ and
$s^{**}=\cos\theta^{**}$ are the values for slipping toward the left
and toward the right respectively, while $s_r=\cos\theta_r=\,^2/_3$
corresponds to the sign inversion of $F$ in the case of the hinged
pencil discussed in the Section\myref{rotating}. The critical value
$\overline{\mu}_s$ of the friction coefficient, beyond which no
left-slipping is possible, can moreover be deduced as the maximum
value of $R(s)$ by requiring that $R\,'(s)=0$: a little algebra
would show indeed that the maximum of $R(s)$ is attained at
$\overline{s}=\,^9/_{11}$, namely at
$\overline{\theta}=\arccos\,^9/_{11}\simeq0.195\,\pi$ corresponding
to the following critical value of the static coefficient of
friction
\begin{equation}\label{critcoeff}
    \overline{\mu}_s^2=R\left(\,^9/_{11}\right)=2^{-13}3^25^3\simeq0.1373\qquad\quad
    \overline{\mu}_s=\frac{15}{64}\sqrt{\frac{5}{2}}\simeq0.3706
\end{equation}
The values of the slipping angles $\theta^*,\theta^{**}$ can finally
be deduced by numerically solving the equation\refeq{criteq}: it is
easy to show for instance that with the values of $\mu_s$ used in
the Figures\myref{critical} and\myref{critical2} we would have
\begin{eqnarray*}
  \mu_s=0.20<\overline{\mu}_s\simeq0.37 && \cos\theta^*\simeq0.964\;\qquad\theta^*\simeq0.086\,\pi \\
  \mu_s=0.60>\overline{\mu}_s\simeq0.37 && \cos\theta^{**}\simeq0.609\qquad\theta^{**}\simeq0.292\,\pi
\end{eqnarray*}
It is also possible to see by direct calculation that at the
critical friction coefficient $\overline{\mu}_s$ of\refeq{critcoeff}
the equation\refeq{Req} in $[0,1]$ also has its smallest solution in
\begin{equation*}
    \overline{\overline{s}}=\frac{48\sqrt{14}-35}{231}
\end{equation*}
corresponding to the the largest solution of\refeq{thetaeq1}
\begin{equation*}
    \overline{\overline{\theta}}=\arccos\frac{48\sqrt{14}-35}{231}\simeq0.285\,\pi
\end{equation*}
By summarizing: the limiting angles are
\begin{equation*}
    0<\overline{\theta}<\theta_r<\overline{\overline{\theta}}<\,^\pi/_2
    \qquad\left\{
            \begin{array}{l}
              \overline{\theta}=\arccos\,^9/_{11}\simeq0.195\,\pi \\
              \theta_r=\arccos\,^2/_3\simeq0.268\,\pi \\
              \overline{\overline{\theta}}=\arccos\frac{48\sqrt{14}-35}{231}\simeq0.285\,\pi
            \end{array}
          \right.
\end{equation*}
and when $\mu_s<\overline{\mu}_s$ the pencil tip slips leftwards
past an angle $\theta^*\le\overline{\theta}$, while if
$\mu_s>\overline{\mu}_s$ a rightward sliding starts only later
beyond an angle $\theta^{**}\ge\overline{\overline{\theta}}$: the
particular values of $\theta^*$ and $\theta^{**}$ depend on $\mu_s$
and can be calculated numerically from the equation\refeq{criteq}.
It also goes without saying that $\theta^*$ grows from $0$ to
$\overline{\theta}$ when $\mu_s$ grows from $0$ to
$\overline{\mu}_s$, while subsequently $\theta^{**}$ starts growing
from $\overline{\overline{\theta}}>\overline{\theta}$ when $\mu_s$
exceeds $\overline{\mu}_s$: no slipping angle (either $\theta^*$ or
$\theta^{**}$) can be found instead in the interval
$[\overline{\theta},\overline{\overline{\theta}}\,]$, namely between
$\arccos\,^9/_{11}\simeq0.195\,\pi$ and
$\arccos\frac{48\sqrt{14}-35}{231}\simeq0.285\,\pi$

Beyond these slipping angles, either $\theta^*$ or $\theta^{**}$,
the pencil dynamics is rather different and we will study in some
detail only the case $\mu_s<\overline{\mu}_s$ with a leftward
slipping beyond $\theta^*$ represented in the Figure\myref{pencil4}:
the case $\mu_s>\overline{\mu}_s$ with a rightward slipping beyond
$\theta^{**}$ is not really different ad its discussion -- combining
elements of the following treatment and of the case of
Section\myref{step} -- is left to the interested reader. When
$\mu_s<\overline{\mu}_s$ we already know that the leftward sliding
of the pen tip begins past an angle $\theta^*=\arccos s^*$ where
$s^*$ is the largest solution of the equation\refeq{criteq} in
$[0,1]$. We also know that this happens at a time $t^*$ that can be
calculated from\refeq{theta} with $\theta=\theta^*$ and in fact
depends on the initial conditions: we recall from the discussion of
the Section\myref{rotating} that in fact $t^*$ diverges when we
choose the zero initial condition $\epsilon\to0^+$, but also that
this is not an insurmountable hindrance if we leave aside the
complete chronological equations and focus instead on the trajectory
shape.

In order to analyze the movement in the intervals $t^*\le t\le T$
and $\theta^*\le\theta\le\,^\pi/_2$ we recall first that the
coordinates of the system of Figure\myref{pencil4} still satisfy the
relations\refeq{coord3} where however we now have $z=0$ for
$0\le\theta\le\theta^*$, and $z\le0$ for
$\theta^*\le\theta\le\,^\pi/_2$. If moreover $\mu_\kappa$ is the
kinetic coefficient of friction between the pencil and the rough
surface, the Newton equations of motion now are
\begin{equation}\label{newt4}
    m\ddot{x}=\mu_\kappa N\qquad\quad m\ddot{y}=N-mg\quad\qquad
    I_{CM}\ddot{\theta}=N\frac{L}{2}\sin\theta
\end{equation}
with $I_{CM}=\,^{mL^2}/_{12}$: these equations coincide with
the\refeq{newt3} of Section\myref{step} but for the first one that
now accounts for the kinetic friction force. Therefore the second
equation\refeq{accel} and the equations\refeq{accel2}
and\refeq{reactN} still hold and hence we can deduce the
equation\refeq{firstint} again as we did in the Sections\myref{rail}
and\myref{step}: here however, to find the integration constant $c$,
we must impose new conditions at $t=t^*$. We have indeed first
from\refeq{omega0} that
\begin{equation*}
    \theta(t^*)=\theta^*\qquad\quad\dot{\theta}(t^*)=\omega^*=\omega(t^*)=\omp\sqrt{1-\cos\theta^*}\qquad\quad\omp=\sqrt{\frac{3g}{L}}
\end{equation*}
and then that
\begin{align*}
    &x(t^*)=\frac{L}{2}\sin\theta^*
      &&\dot{x}(t^*)=\frac{L}{2}\omp\cos\theta^*\sqrt{1-\cos\theta^*}\\
    &y(t^*)=\frac{L}{2}\cos\theta^*
      &&\dot{y}(t^*)=-\frac{L}{2}\omp\sin\theta^*\sqrt{1-\cos\theta^*}\\
    &z(t^*)=0
      &&\dot{z}(t^*)=0
\end{align*}
We are therefore able to calculate $c$ and after a little algebra we
find
\begin{equation}\label{angvel4}
    \dot{\theta}^2=\frac{4\,\omp^2}{9}\,\frac{9-\frac{27}{4}\cos^2\theta^*(1-\cos\theta^*)-9\cos\theta}{4-3\cos^2\theta}
\end{equation}
that replaces\refeq{angvel3} with its corresponding time equation
which is now
\begin{equation}\label{theta4}
    \int_{\theta^*}^\theta\sqrt{\frac{4-3\cos^2\phi}{9-\frac{27}{4}\cos^2\theta^*(1-\cos\theta^*)-9\cos\phi}}\,d\phi=\frac{2\,\omp}{3}\,
    (t-t^*)
\end{equation}
This integral can be numerically evaluated to calculate the time $t$
needed to reach an angle $\theta\in[\theta^*,\,^\pi/_2]$: for
instance, if we take
$\cos\theta^*=0.95>\,^9/_{11}=\cos\overline{\theta}$ (that
corresponds to $\mu_s\simeq0.233<0.371=\overline{\mu}_s$), the time
$T$ when the pencil hits the floor now becomes
\begin{equation}\label{impacT2}
    T=t^*+\frac{3}{2\omp}\int_{\theta^*}^{\,^\pi/_2}\sqrt{\frac{4-3\cos^2\phi}{9-\frac{27}{4}\cos^2\theta^*(1-\cos\theta^*)-9\cos\phi}}\,d\phi
    \simeq t^*+\frac{2.029}{\omp}
\end{equation}
where $t^*$ comes from\refeq{theta} choosing a small initial
condition $\epsilon>0$. As for the reaction force $N$ on the other
hand, from\refeq{newt4} (namely\refeq{accel2} and\refeq{reactN} as
in the Sections\myref{rail} and\myref{step}) and from\refeq{angvel4}
we have now
\begin{figure}
\begin{center}
\includegraphics[width=11cm]{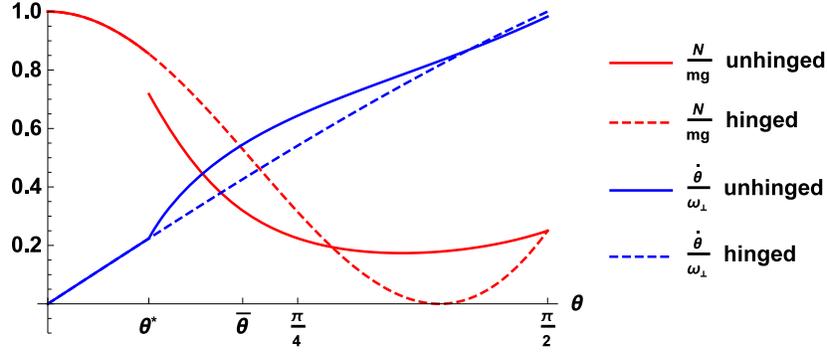}
\caption{Dimensionless reaction force $\,^N/_{mg}$ and angular
velocity $\,^{\dot{\theta}}/_{\omp}$ (continuous lines) on a rough
surface with $\mu_s<\overline{\mu}_s$, compared with the same
quantities in the case of the hinged pencil (dashed lines): the
values coincide when $\theta<\theta^*$. Here we took
$\cos\theta^*=0.95$ corresponding to
$\mu_s\simeq0.233$}\label{Nrough}
\end{center}
\end{figure}
\begin{equation}\label{N4}
    N=mg\,\frac{3\cos^2\theta-\big[6-\frac{9}{2}(1-\cos\theta^*)\cos^2\theta^*\big]\cos\theta+4}{(4-3\cos^2\theta)^2}
    \quad\qquad\theta^*\le\theta\le\,^\pi/_2
\end{equation}
while for $0\le\theta\le\theta^*$ it takes the same values of the
hinged case of Section\myref{rotating}. The plot of $\,^N/_{mg}$ in
the Figure\myref{Nrough} shows in particular that $N$ is
discontinuous at $\theta^*$ signaling the transition from the static
to the kinetic friction. In the same Figure\myref{Nrough} also the
dimensionless angular velocity $\,^{\dot{\theta}}/_{\omp}$ is
displayed in the same intervals.

\begin{figure}
\begin{center}
\includegraphics[width=12cm]{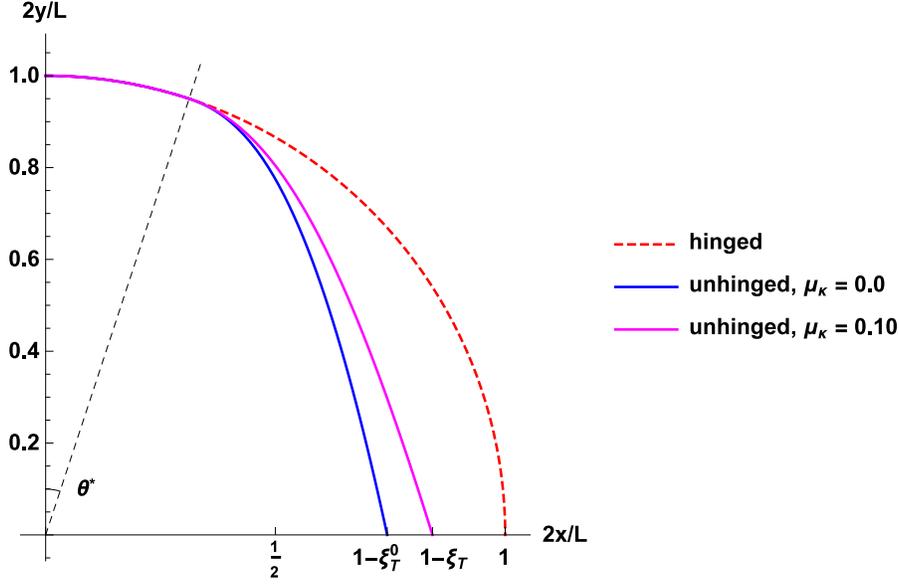}
\caption{Dimensionless \emph{CM} $xy$-trajectory: it coincides with
the circular path of the hinged pencil for $\theta\le\theta^*$, but
as soon as $\theta\ge\theta^*$ it follows different flight according
to the kinetic coefficient of friction: the parametric equations
are\refeq{param} together with\refeq{zeta2} to calculate
$\zeta(s)$.}\label{landing1}
\end{center}
\end{figure}

We come finally to give some detail about the \emph{CM} trajectory
and the position $z$ of the tip, but at variance with the discussion
of the Section\myref{step}, $\dot{x}(t)$ no longer is a constant as
in\refeq{CMvx} since we must now take into account the kinetic
friction force in the first equation\refeq{newt4}. A quest for a
simple chronological equation $x(t)$, however, would still be doomed
because of the rather involuted form\refeq{N4} of $N$. We can
nevertheless gain some insight into the trajectories by looking
again to our quantities rather as functions of the angle $\theta$,
as we already did in the previous sections. While apparently for
$0\le\theta\le\theta^*$ it is $z=0$ and the \emph{CM} follows a
circular path of radius $\,^L/_2$ around the origin, as soon as
$\theta>\theta^*$ it will follow a path of parametric
equations\refeq{param} with $\zeta(s)=0$ for
$s=\cos\theta\in[s^*,1]$ ($s^*=\cos\theta^*$): in order to complete
the trajectory we are therefore left just with the task of
calculating $\zeta(s)$ for $s\in[0,s^*]$. In order to do that we
first remark that from\refeq{coord3} we have
\begin{equation*}
    \dot{z}=\dot{x}-\frac{L}{2}\dot{\theta}\cos\theta
\end{equation*}
On the other hand, within the notations of the Section\myref{step}
with $s=\cos\theta$, it is
\begin{figure}
\begin{center}
\includegraphics[width=11cm]{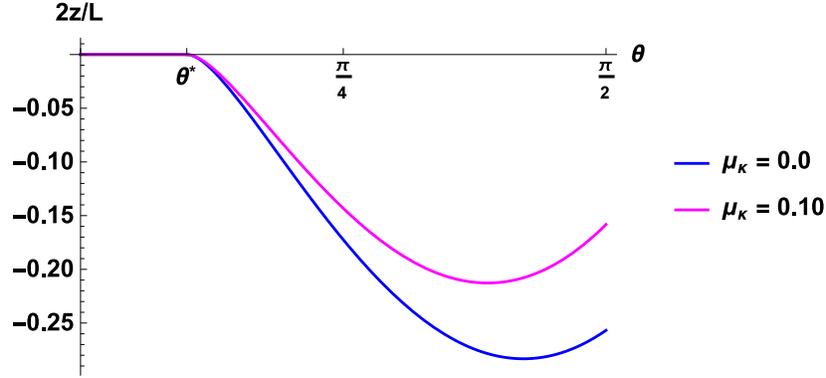}
\caption{Dimensionless position $\,^{2z(\theta)}/_L$ of the pencil
tip laid on a rough table for two different kinetic friction
coefficients $\mu_k$, as a function of $\theta$: as long as
$\theta\le\theta^*$ it is $\,^{2z(\theta)}/_L=0$, but when
$\theta\ge\theta^*$ the function $z(\theta)=\zeta(\cos\theta)$
should be calculated from\refeq{zeta2}: its value is now in the
negative. Here again we have chosen $\cos\theta^*=0.95$.}\label{z1}
\end{center}
\end{figure}
\begin{equation*}
    \dot{z}=\zeta\,'\dot{s}=-\zeta\,'\dot{\theta}\sin\theta=-\zeta\,'\dot{\theta}(s)\sqrt{1-s^2}
\end{equation*}
so that, defining a function $v(s)$ such that
$\dot{x}(t)=v\big(s(t)\big)$, we get
\begin{equation}\label{zeta1}
    \zeta\,'(s)=-\frac{v(s)}{\dot{\theta}(s)\sqrt{1-s^2}}+\frac{L}{2}\,\frac{s}{\sqrt{1-s^2}}
    \qquad\quad\zeta(s^*)=0
\end{equation}
We see moreover from the definitions that
\begin{equation*}
    \ddot{x}=v'\dot{s}=-v'(s)\dot{\theta}(s)\sqrt{1-s^2}
\end{equation*}
and hence the first dynamical equation\refeq{newt4} becomes
\begin{equation*}
    v'(s)=-\frac{\mu_\kappa}{m}\,\frac{N(s)}{\dot{\theta}(s)\sqrt{1-s^2}}\qquad\quad
    v(s^*)=v^*=\frac{L\,\omp}{2} s^*\sqrt{1-s^*}
\end{equation*}
to wit
\begin{equation}\label{v}
    v(s)=v^*+\frac{\mu_\kappa}{m}\int_s^{s^*}\frac{N(r)}{\dot{\theta}(r)\sqrt{1-r^2}}\,dr
\end{equation}
By assembling\refeq{zeta1} and\refeq{v} we finally have
\begin{equation}\label{zeta2}
    \zeta(s)=\int_s^{s^*}\left[\frac{1}{\dot{\theta}(q)\sqrt{1-q^2}}
    \left(v^*+\frac{\mu_\kappa}{m}\int_q^{s^*}\frac{N(r)}{\dot{\theta}(r)\sqrt{1-r^2}}\,dr\right)
    -\frac{L}{2}\,\frac{q}{\sqrt{1-q^2}}\right]dq
\end{equation}
where, with $\beta^*=s^*\sqrt{1-s^*}$, it is understood
from\refeq{angvel4} and\refeq{N4} that
\begin{equation*}
    \dot{\theta}(s)=\omp\sqrt{\frac{4-3\beta^{*\,2}-4s}{4-3s^2}}\qquad\quad
    N(s)=mg\frac{6s^2-3(4-3\beta^{*\,2})s+8}{2(4-3s^2)}
\end{equation*}
The integral\refeq{zeta2} can be calculated numerically and lends
again the possibility of plotting both $z(\theta)=\zeta(\cos\theta)$
(Figure\myref{z1}), and the trajectory parametric
equations\refeq{param} together with\refeq{zeta2}
(Figure\myref{landing1}) where it is understood that $\zeta(s)=0$
when $s^*\le s\le1$. In both the plots we have chosen
$\theta^*=\arccos 0.95$ (corresponding to the static coefficient of
friction $\mu_s\simeq0.233$), and two possible values for the
kinetic coefficient of friction: the limiting value
$\mu_\kappa^0=0.0$ and $\mu_\kappa=0.10\,$. Remark that now, at
variance with what we have found in the similar discussion of the
Section\myref{step}, $z(s)$ takes negative values for $0\le s\le
s^*$ accounting for the fact that the pencil tip slides leftward.
From\refeq{zeta2} we can also calculate the point $x_T$ where the
pencil \emph{CM} hits the floor at the time $T$: since it is
$\theta(T)=\,^\pi/_2$, namely $s(T)=0$ we will have
\begin{equation*}
    x_T=z_T+\frac{L}{2}=\zeta(0)+\frac{L}{2}=\frac{L}{2}(1-\xi_T)\qquad\quad\xi_T=-\frac{2\zeta(0)}{L}\ge0
\end{equation*}
where, with $\theta^*=\arccos 0.95$, it is
\begin{equation*}
                                 \left\{
                                      \begin{array}{ll}
                                        \xi_T^0\simeq0.257, & \quad\hbox{for $\mu_\kappa^0=0.0$} \\
                                        \xi_T\simeq0.158, & \quad\hbox{for $\mu_\kappa=0.10$}
                                      \end{array}
                                    \right.
\end{equation*}

\section{Epilogue}\label{concl}

In this paper we have given an elementary treatment of a mechanical
case study: the dynamics of a pencil with a tip laid on a rough
table and set free to fall under the action of gravity. Despite its
seeming modesty and lack of pretention we have shown that a
discussion of this simple problem still conceals many details of
(maybe) unexpected -- but never unsurmountable -- intricacy that may
turn out to be pedagogically edifying. Along our exploration we also
had the occasion to point out a few small results of a broader
scope, as for instance some critical values of the sliding angles
and of the static coefficients of friction. We hope that this
\emph{Divertimento} could eventually prove to be both profitable and
entertaining for all those willing to stop for a while to listen at
it


\begin{thebibliography}{99}

\bibitem{frank}\emph{G.F.\ Franklin, J.D.\ Powell and A. Emami-Naeini}, \textsc{Feedback Control of Dynamic Systems}
(Pearson, Boston \textbf{2015})

\bibitem{liber}\emph{D.\ Liberzon}, \textsc{Switching in Systems and
Control} (Birkh\"auser, Boston, \textbf{2003})

\bibitem{www}
\textsf{www.grc.nasa.gov/WWW/k-12/VirtualAero/BottleRocket/airplane/rktstab.html}
\\
\textsf{www2.math.ou.edu/{\footnotesize$\sim$}npetrov/joe-report.pdf }\\
\textsf{robotics.ee.uwa.edu.au/theses/2003-Balance-Ooi.pdf}

\bibitem{grad}\emph{I.S.\ Gradshteyn and I.M.\ Ryzhik}, \textsc{Table of
Integrals, Series and Products} (Academic Press, Burlington
\textbf{2007})

\end{thebibliography}
\end{document}